%% file: main.tex
\documentclass{article}

\PassOptionsToPackage{numbers, compress}{natbib}
\usepackage[final]{neurips_2022_ml4ps}

\usepackage[utf8]{inputenc} 
\usepackage[T1]{fontenc}    
\usepackage{url}            
\usepackage{booktabs}       
\usepackage{amsfonts}       
\usepackage{nicefrac}       
\usepackage{microtype}      
\usepackage{xcolor}         

\usepackage{bm}             
\usepackage{multirow}
\usepackage{amsmath}
\usepackage{amssymb}
\usepackage{subcaption}
\usepackage[pdftex]{graphicx}
\usepackage[symbol]{footmisc}

\newcommand{\blank}{\mathrel{\;\cdot\;}}

\title{Physics-Informed Convolutional Neural Networks for Corruption Removal on Dynamical Systems}

\author{
  Daniel Kelshaw \\
  Department of Aeronautics \\
  Imperial College London \\
  \texttt{djk21@imperial.ac.uk} \\
  \And
  Luca Magri \\
  Department of Aeronautics \\
  Imperial College London, \\
  Alan Turing Institute \\
  \texttt{l.magri@imperial.ac.uk} \\
}

\begin{document}

\maketitle

\begin{abstract}
    Measurements on dynamical systems, experimental or otherwise, are often subjected to inaccuracies
    capable of introducing corruption; removal of which is a problem of fundamental importance in the 
    physical sciences. In this work we propose physics-informed convolutional neural networks for 
    stationary corruption removal, providing the means to extract physical solutions from data, given 
    access to partial ground-truth observations at collocation points. We showcase the methodology for
    2D incompressible Navier-Stokes equations in the chaotic-turbulent flow regime, demonstrating 
    robustness to modality and magnitude of corruption.
\end{abstract}

\begin{figure}[ht]
    \centering
    \includegraphics[width=4.5in, height=1.3in]{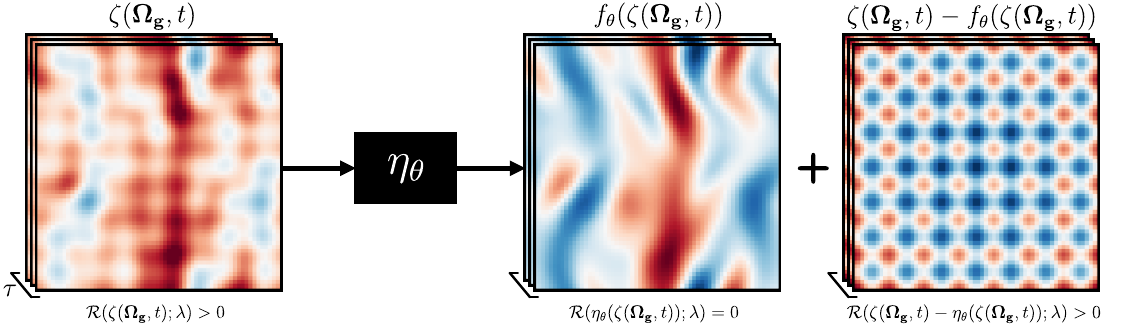}
    \caption{Diagrammatic overview of the corruption removal task.}
    \label{fig:overview}
\end{figure}

\input{sections/01_introduction}
\input{sections/02_methodology}
\input{sections/03_results}
\input{sections/04_conclusion}

\acksection{
D. Kelshaw. and L. Magri. acknowledge support from the UK EPSRC. 
L. Magri gratefully acknowledges financial support from the ERC Starting Grant PhyCo 949388.
}


\bibliography{references}

\end{document}

%% file: sections/01_introduction.tex
\section{Introduction}
In the physical sciences, corrupted data can arise for a number of reasons: faulty experimental sensors, or low-fidelity numerical results to name
a few. The detection and removal of this corruption would allow for the extraction of the underlying true-state, ensuring that the governing equations
are satisfied. In practice, verifying that given observations are characteristic of a dynamical system is more straightforward than determining the 
solution itself. For observations on dynamical systems we evaluate the residual, giving an indication of the fidelity of the observations.

Corruption removal typically considers noise-removal, and flow-reconstruction \cite{Brunton2020}. Noise-removal constitutes methods which remove 
small, stochastic variations from the true-state, with examples including: filtering methods \cite{Kumar2017}, proper orthogonal decomposition 
\cite{Raiola2015, Mendez2017}, and autoencoders \cite{Vincent2008}. Flow-reconstruction considers inference of a solution over the domain, given only 
partial information; a popular example being that of super-resolution. Convolutional neural networks (CNNs) are prominent in the domain of flow 
reconstruction due to their ability to exploit spatial correlations \cite{Dong2014, Shi2016}.

In the absence of ground-truth labels, it is possible to impose prior knowledge by regularising predictions with respect to known governing equations
\cite{Lagaris1998}. Physics-informed CNNs have been used for solving PDEs \cite{Gao2021, Ren2022}, basic super-resolution \cite{GaoSR2021}, and 
surrogate modelling without access to labelled data \cite{Zhao2021}. This regularisation method differs from the standard physics-informed neural 
network (PINN) approach which strives to exploit the automatic-differentiation capabilities of neural networks to constrain gradients of the network \cite{Raissi2019}.

Our work considers a form of physics-informed CNNs for flow reconstruction. We provide a method for removing arbitrary, stationary corruption from
observations on dynamical systems, given access to only partial ground-truth information at select collocation points. We showcase corruption removal
for the 2D incompressible Navier-Stokes equations in the chaotic-turbulent regime, highlighting the ability to accurately recover the true-state, independent 
of modality or magnitude of corruption.

%% file: sections/02_methodology.tex
\section{Methodology} \label{sec:methodology}

We consider corruption removal from observations on dynamical systems of the form

\begin{equation} \label{eqn:dynamical_system}
    \partial_t \bm{u} - \mathcal{N}(\bm{u}; \lambda) = 0
        \qquad
        \text{with}
            \quad \bm{u} = u(\bm{x}, t),
            \quad \bm{x} \in \Omega \subset \mathbb{R}^{n}, 
            \quad t \in [0, T] \subset \mathbb{R}^{+},
\end{equation}

where $u: \Omega \times [0, T] \rightarrow \mathbb{R}^{n}$, $\mathcal{N}$ is a sufficiently smooth differential operator, and $\lambda$ 
are the physical parameters of the system. We define the residual of the system as the left-hand side of equation (\ref{eqn:dynamical_system})

\begin{equation}
    \mathcal{R}(\bm{u}; \lambda) \triangleq \partial_t \bm{u} - \mathcal{N}(\bm{u}; \lambda),
\end{equation}

such that $\mathcal{R}(\bm{u}; \lambda) = 0$ when $u(\bm{x}, t)$ is a solution to the partial differential equation (PDE). We consider
corrupted observations such that $\mathcal{R}(\zeta(\bm{x}, t); \lambda) > 0$, where $\zeta(\bm{x}, t)$ is the linear combination of the 
true state $u(\bm{x}, t)$ and the stationary corruption $\phi(\bm{x})$, precisely $\zeta(\bm{x}, t) = u(\bm{x}, t) + \phi(\bm{x}).$

Given the corrupted state, we aim to recover the underlying, physical solution to the governing equations: the true-state. Mathematically,
we denote this process by the mapping

\begin{equation}
    \eta_{\theta}: \zeta(\bm{\Omega_{g}}, t) \rightarrow u(\bm{\Omega_{g}}, t),
\end{equation}

where the domain $\Omega$ is discretised on a uniform, structured grid $\bm{\Omega_{g}} \subset \mathbb{R}^{N^n}$. We further discretise
the time domain, providing $\mathcal{T} = \{ t_i \in [0, T] \}_{i=0}^{N_t}$ for $N_t$ samples in time. Approximating the mapping $\eta_{\theta}$
as a CNN, we optimise weights of the network to minimise the loss

\begin{equation}
    \mathcal{L}_{\theta} = \alpha \left( \mathcal{L}_{\partial \Omega} + \mathcal{L}_{\phi} \right) + \mathcal{L}_{\mathcal{R}},
\end{equation}

where $\alpha$ is a fixed, empirical weighting factor. We define each of the loss terms as

\begin{align}
    \mathcal{L}_{\mathcal{R}} &= \frac{1}{\lvert \mathcal{T} \rvert} \sum_{t \in \mathcal{T}} \big\lVert
        \mathcal{R}\big(\eta_{\theta}(\zeta(\bm{\Omega_{g}}, t)); \lambda \big)
   \big\rVert_{\bm{\Omega_{g}}}^{2}, \\
   \mathcal{L}_{\partial \Omega} &= \frac{1}{\lvert \mathcal{T} \rvert} \sum_{t \in \mathcal{T}} \big\lVert
        \eta_{\theta}(\zeta(\partial \bm{\Omega_{g}}, t)) - u(\partial \bm{\Omega_{g}}, t)
    \big\rVert_{\partial \bm{\Omega_{g}}}^{2}, \\
   \mathcal{L}_{\phi} &= \frac{1}{\lvert \mathcal{T} \rvert} \sum_{t \in \mathcal{T}} \big\lVert
        \partial_t \big[ \zeta(\bm{\Omega_{g}}, t) - \eta_{\theta}(\zeta(\bm{\Omega_{g}}, t)) \big]
   \big\rVert_{\bm{\Omega_{g}}}^{2},
\end{align}

where $\partial \bm{\Omega_{g}}$ denotes boundary points of the grid, and $\lVert \blank \rVert_{\Omega}$ is the $\ell^2$-norm over the 
given domain. We regularise network predictions with the residual-based loss $\mathcal{L}_{\mathcal{R}}$, imposing prior-knowledge of
the system to promote network realisations which conform to the governing equations. The data-driven boundary loss $\mathcal{L}_{\partial \Omega}$
serves to minimise the error between predictions and known measurements; when used in conjunction with the residual-loss, this effectively 
allows us to condition the underlying physics on the observations. For this case, we consider access to ground-truth measurements on the boundary
$\partial \Omega$. Inclusion of the corruption-based loss $\mathcal{L}_{\phi}$ embeds our assumption of stationary corruption, driving predictions
away from the trivial solution, $u(\bm{\Omega_{g}}, t) = 0$.\footnote[2]{All code is available on GitHub: \url{https://github.com/magriLab/PICR}}

%% file: sections/03_results.tex
\newpage
\section{Results} \label{sec:results}

We consider the corruption removal for measurements on the Kolmogorov flow, an instance of the 2D incompressible Navier-Stokes equations. 
The flow is defined on the domain $\Omega \in [0, 2\pi) \subset \mathbb{R}^{2}$, imposing periodic boundary conditions on $\partial \Omega$,
and subjected to stationary, sinusoidal forcing $g(\bm{x})$. The chaotic-turbulent nature of the Kolmogorov flow provides a 
challenging case study, introducing complex non-linear dynamics and turbulent structures across a range of spatial frequencies \cite{Fylladitakis2018}. 
The standard continuity and momentum equations are given as
\vspace{0.5em}
\begin{align} \label{eqn:kolmogorov}
\begin{split}
    \nabla \cdot \bm{u} &= 0 \\
    \partial_t \bm{u} + \bm{u} \cdot \nabla \bm{u} &= - \nabla p + \nu \Delta \bm{u} + g(\bm{x}),
\end{split}
\end{align}

where $p, \nu$ denote the scalar pressure field and kinematic viscosity respectively. We take $\nu = \nicefrac{1}{34}$ to ensure 
chaotic-turbulent dynamics, and prescribe the forcing model $g(\bm{x}) = [\sin(4\bm{x}_{2}), 0]^{\top}$. 

\subsection{Pseudospectral Discretisation} \label{sec:spectral}
We provide a differentiable pseudospectral discretisation for the differential operator, defining operations on the spectral grid
$\bm{\hat{\Omega}_{k}} \in \mathbb{Z}^{61 \times 61}$, enabling backpropagation for the residual-based loss $\mathcal{L}_{\mathcal{R}}$. 
By eliminating the pressure term, our discretisation handles the continuity constraint implicitly, affording us the ability to neglect
the constraint in the loss \cite{Canuto1988}. We produce a solution by time-integration of the dynamical system with 
the forward-Euler method, taking a time-step $\Delta t = 0.005$ to ensure numerical stability according to the Courant-Friedrichs-Levy 
condition.

Due to the spectral discretisation, we evaluate the residual-based loss $\mathcal{L}_{\mathcal{R}}$ in the Fourier domain on the grid 
$\bm{\hat{\Omega}_{k}}$, offering two distinct advantages over the standard finite-difference approach: spectral accuracy, and implicit handling
of periodic boundary conditions. We employ a form of time-windowed batching, considering the evaluation of the residual-based loss over 
$\tau \geq 2$ consecutive time-steps.

\subsection{Parameterised Corruption}
We conduct experiments varying the frequency $f$, and relative magnitude $\mathcal{M}$ of the corruption. For a scaling magnitude 
$M = \mathcal{M} \cdot \max{\left(u(\bm{\Omega_{g}},\blank)\right)}$, we define the corruption field as
\vspace{0.3em}
\begin{equation}
    \phi(\bm{x}; f, M) = \Gamma \left(20 + \sum_{i=1}^{2} \left[ (\bm{x}_{i} - \pi)^{2} - 10 \cos(f \cdot (\bm{x}_{i} - \pi))  \right]; M \right),
\end{equation}
\vspace{0.3em}
where $\Gamma(\; \bm{\cdot} \; ; M)$ is a linear operator that re-scales the vector in the range $[0, M]$, \cite{Rastrigin1974}.
\vspace{0.01em}
\begin{figure*}[htb]
    \centering
    \includegraphics[width=\linewidth]{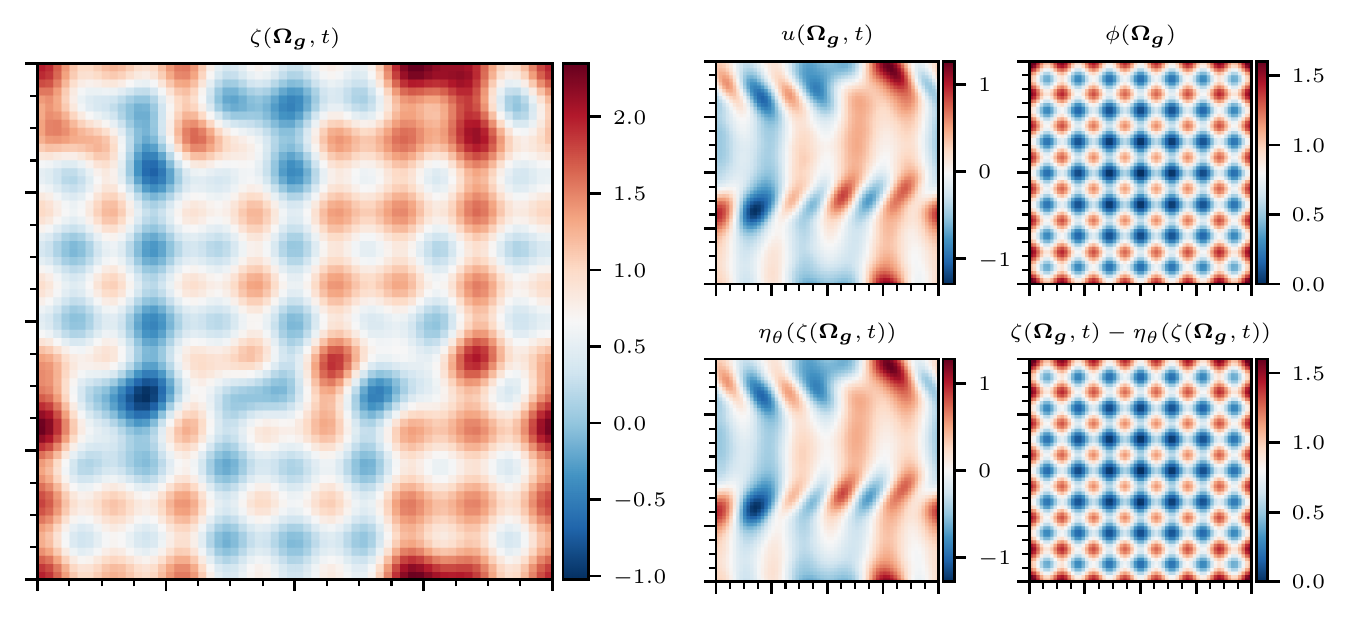}
    \caption{Corruption removal results for the Kolmogorov flow $[f=7, \mathcal{M}=0.5]$.}
    \label{fig:kolmogorov_results}
\end{figure*}

\subsection{Numerical Experiments} \label{sec:numerical_experiments}
The Kolmogorov flow is solved in the Fourier domain, as per §\ref{sec:spectral}, before generating data on the grid 
$\bm{\Omega_{g}} \in \mathbb{R}^{64 \times 64}$. A total of $1024$ time-windows are used for training, with a further $256$ for validation, taking 
$\tau = 2$ in each instance. We employ a standard U-Net \cite{Ronneberger2015} architecture and train with the Adam optimizer, taking a learning 
rate of $3\times10^{-4}$ for $5000$ epochs, fixing computational resource for each experiment. We impose a fixed weighting of $\alpha = 1\times10^{3}$ 
for the empirical weighting term.\footnote[3]{All experiments were run on a single NVIDIA Quadro RTX 8000.} 

In Figure \ref{fig:kolmogorov_results} we demonstrate results for $f = 7, \mathcal{M} = 0.5$, showcasing the ability to separate the true-state from
corruption. We note that the residual of the corruption is not equal to the sum of the residuals of it's constituents, highlighting the challenge of
non-linearity. We further assess the robustness of the methodology to both modality and magnitude of corruption by running two studies: 

\begin{equation*}
\begin{aligned}[c]
    &\mathit{(i)} \\
    &\mathit{(ii)}
\end{aligned}
\quad
\begin{aligned}[c]
    &\mathcal{M} = 0.5, \\
    &f = 3,
\end{aligned}
\quad
\begin{aligned}[c]
    f &\in \{ 1, 3, 5, 7, 9 \}; \\
    \mathcal{M} &\in \{ 0.01, 0.1, 0.25, 0.5, 1.0 \}.
\end{aligned}
\end{equation*}

\begin{figure}[htb]
    \centering
    \includegraphics[width=\linewidth]{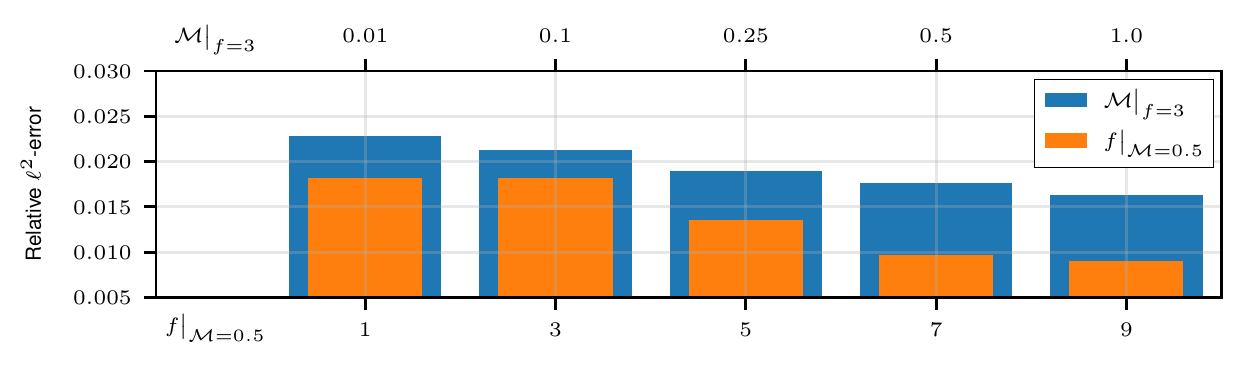}
    \caption{
        Comparison of relative $\ell^2$-error for parameterised corruption frequency and magnitude. 
    }
    \label{fig:param_runs}
\end{figure}

Results can be seen in Figure \ref{fig:param_runs}, where performance for all experiments is measured by the relative $\ell^2$-error between the 
predicted-state $\eta_{\theta}(\zeta(\bm{\Omega}_{g}, t))$, and the true-state $u(\bm{\Omega_{g}}, t)$. Results denote the mean over five repeats
with random seeds, the standard deviation not exceeding $8.169 \times 10^{-4}$.

We find that performance is largely agnostic to both multi-modality and magnitude of corruption, with error decreasing slightly with increasing $f, \mathcal{M}$,
as demonstrated in Figure \ref{fig:param_runs}. Quantitatively, we observe excellent robustness of the methodology, with relative $\ell^2$-error not exceeding
$2.284 \times 10^{-2}$. 

%% file: sections/04_conclusion.tex
\section{Conclusion} \label{sec:conclusion}

In this work we introduce a methodology for physics-informed removal of arbitrary, stationary corruption from observations on dynamical
systems, given access to only partial ground-truth information. We demonstrate results for corrupted observations on the Kolmogorov flow,
using ground-truth observations on the boundaries to condition the physics. Experiments indicate that performance is agnostic to both
magnitude and modality of corruption, with relative $\ell^2$-error remaining consistently low, as demonstrated in Figure~\ref{fig:param_runs}.
Our results in Figure \ref{fig:kolmogorov_results} qualitatively showcase the performance of the model, clearly separating the corrupted
field into its component parts: true-state, and corruption. This work opens opportunities for accurate flow reconstruction from corrupted
state measurements with access to partial ground-truth information, potentially allowing for correcting sensor measurements and low-fidelity
simulation results - problems of fundamental importance in the physical sciences.

We note the limitations of our work: the assumption of stationary corruption, and the requirement of ground-truth observations to condition
the physics; providing considerations for future work.